\documentclass[twocolumn,prd,nofootinbib,superscriptaddress,longbibliography]{revtex4} %
\newcommand\ForInternalReference[1]{}
\newcommand\SkipForEarlyCirculation[1]{}

\newcommand\SkipPP[1]{}
\usepackage[colorlinks=true,citecolor=blue,urlcolor=blue]{hyperref}
\usepackage{verbatim}
\usepackage{graphicx}
\usepackage{dcolumn}
\usepackage{bm}
\usepackage{color}
\usepackage{xcolor}
\usepackage{url}
\usepackage{float}
\usepackage{multirow}
\usepackage{adjustbox}
\usepackage[normalem]{ulem}
\usepackage{bm}
\usepackage{placeins}
\usepackage{afterpage}
\usepackage{amsmath,amssymb}
\usepackage{acronym}
\usepackage{booktabs}      
\usepackage{tabularx}      
\usepackage{array}         

\newcommand\optional[1]{}
\acrodef{NR}[NR]{Numerical Relativity}

\newcommand\lnLmarg{ \ln{\cal L}_{\rm marg}{}}

\usepackage{color}
\definecolor{amber}{rgb}{1.0, 0.75, 0.0}
\definecolor{orange}{rgb}{1.0, 0.5, 0.0}
\definecolor{amaranth}{rgb}{0.9, 0.17, 0.31}

\graphicspath{{./figures/}}
\newcommand{\mc}{{\cal M}}

\def\ltsima{$\; \buildrel < \over \sim \;$}
\def\simlt{\lower.5ex\hbox{\ltsima}}
\def\gtsima{$\; \buildrel > \over \sim \;$}
\def\simgt{\lower.5ex\hbox{\gtsima}}

\def\eos#1{equation of state#1 (EOS#1)\gdef\eos{EOS}}
\def\QNM#1{quasi-normal mode#1 (QNM#1)\gdef\QNM{QNM}}
\def\ns#1{neutron star#1 (NS#1)\gdef\ns{NS}}
\def\gw#1{gravitational wave#1 (GW#1)\gdef\gw{GW}}
\def\bh#1{black hole#1 (BH#1)\gdef\bh{BH}}
\def\bbh#1{binary black hole#1  (BBH#1)\gdef\bbh{BBH}}
\def\bns#1{binary neutron star#1 (BNS#1)\gdef\bns{BHS}}
\def\bhns#1{black hole - neutron star#1 (BHNS#1)\gdef\bhns{BHNS}}
\def\nsbh#1{neutron star - black hole#1 (NSBH#1)\gdef\nsbh{NSBH}}
\def\nr#1{numerical relativity#1 (NR#1)\gdef\nr{NR}}

\newcommand{\UT}{\affiliation{Center for Gravitational Physics, The University of Texas at Austin, Austin, Texas 78712, USA}}

\begin{document}

\title{GW200105: A detailed study of eccentricity in the neutron star-black hole binary} 
\author{Aasim Jan}
\UT
\author{Bing-Jyun Tsao}
\UT 
\author{Richard O'Shaughnessy}
\affiliation{Center for Computational Relativity and Gravitation, Rochester Institute of Technology, Rochester, New York 14623, USA}
\author{Deirdre Shoemaker}
\UT
\author{Pablo Laguna}
\UT

\begin{abstract}
GW200105\_162426 is the first neutron star–black hole merger to be confidently confirmed through either gravitational-wave or electromagnetic observations. Although initially analyzed after detection, the event has recently gained renewed attention following a study  [Morras \textit{et al.} \href{https://arxiv.org/abs/2503.15393}{\color{blue} arXiv:2503.15393}] that employed a post-Newtonian inspiral-only waveform model and reported strong evidence for orbital eccentricity. In this work, we perform a detailed analysis of GW200105 using state-of-the-art effective-one-body waveform models.  Importantly, we present the first study of this event utilizing a physically complete model that incorporates both orbital eccentricity and spin precession across the full inspiral, merger, and ringdown stages, along with higher-order gravitational wave modes.  Our results support the presence of eccentricity in the signal, with zero eccentricity excluded from the $99\%$ credible interval, but yielding a mass ratio closer to the original LIGO-Virgo-KAGRA analysis, differing from the findings of [Morras \textit{et al.} \href{https://arxiv.org/abs/2503.15393}{\color{blue} arXiv:2503.15393}]. Additionally, similar to a previous eccentric-only analysis [de Lluc Planas \textit{et al.} \href{https://iopscience.iop.org/article/10.3847/1538-4357/ae1d7d}{\color{blue}
Astrophys. J. 995, 47 (2025).}], we observe a multimodal structure in the eccentricity posterior distribution. We conduct targeted investigations to understand the origin of this multimodality and complement our analysis with numerical relativity simulations to examine how the inclusion of eccentricity impacts the merger dynamics.
\end{abstract}
\maketitle

\section{Introduction}
On January 5, 2020, the LIGO-Virgo \cite{Aasi_2015,Acernese_2015} network detected gravitational waves (GWs) from a compact binary system, with the signal consistent with a neutron star–black hole (NSBH) merger \cite{Abbott_2021}. This event, designated GW200105, is particularly significant as it marks the first confident detection of an NSBH system via either gravitational or electromagnetic observations. Alongside a second NSBH event, GW200115 \cite{Abbott_2021}, detected just ten days later, GW200105 provided crucial evidence that neutron stars and black holes can indeed form binaries and coalesce within a Hubble time, an outcome long anticipated by theoretical models but previously unconfirmed by observations \cite{Abbott_2021}.

In its initial analysis, the LIGO-Virgo-KAGRA (LVK) Collaboration estimated the source-frame primary mass of this binary to be $8.9^{+1.1}_{-1.3}$, well above the maximum neutron star mass \cite{PhysRevLett.32.324,PhysRevLett.121.161101,Cromartie_2019,PhysRevD.100.023015,Farr_2020,Fonseca_2021,2021ApJ...908L..28N}, while the source-frame secondary mass was estimated to be $1.9^{+0.2}_{-0.2}$, consistent with the known mass range for neutron stars  
\cite{antoniadis2016millisecondpulsarmassdistribution,2017PhRvL.119p1101A,2018MNRAS.478.1377A}. The analysis reported no evidence of tidal deformation or spin precession. Additionally, it did not address the possibility of orbital eccentricity as the waveform models employed assumed quasicircular orbits. This choice is well justified as binaries formed through isolated stellar evolution 
are expected to circularize before entering the detector frequency band \cite{Stevenson_2017}. Furthermore, full inspiral-merger-ringdown (IMR) waveform models capable of capturing the effects of orbital eccentricity have only recently become available \cite{PhysRevD.101.044049,PhysRevD.105.044035,PhysRevD.110.024031,gamboa2024accuratewaveformseccentricalignedspin,planas2025timedomainphenomenologicalmultipolarwaveforms}.

Since then, GW200105 has attracted renewed attention, particularly regarding the possibility of nonzero orbital eccentricity. A recent analysis \cite{morras2025orbitaleccentricityneutronstar} employing a post-Newtonian (PN) waveform model \cite{Morras_2025} that simultaneously accounts for eccentricity and spin precession found the signal to be consistent with an eccentric binary, estimating an orbital eccentricity of $0.145^{+0.007}_{-0.09}$ at a reference frequency of 20 Hz. This analysis also inferred a detector-frame total mass $M_\text{tot}$ of $13.86^{+0.67}_{-1.69}$ and a mass ratio $q$ of $0.13^{+0.05}_{-0.01}$. These results are consistent with the possibility, suggested by \cite{Fei_2024}, that the event is eccentric. An independent analysis \cite{planas2025eccentricinspiralmergerringdownanalysisneutron} using a complete IMR waveform model incorporating eccentricity (but restricted to aligned spin configurations) \cite{planas2025timedomainphenomenologicalmultipolarwaveforms} also found strong evidence for orbital eccentricity, estimating an eccentricity of $0.12^{+0.02}_{-0.03}$, with $M_\text{tot} = 10.83^{+1.06}_{-0.60}$ and $q = 0.24^{+0.04}_{-0.05}$. Notably, the eccentricity posterior distribution in this analysis exhibited a bimodal structure, in contrast to the unimodal eccentricity distribution found in \cite{morras2025orbitaleccentricityneutronstar}. Another study \cite{kacanja2025eccentricitysignaturesligovirgokagrasbns}, which also employed complete IMR waveform models with eccentricity under aligned-spin assumptions, reported findings consistent with those of \cite{planas2025eccentricinspiralmergerringdownanalysisneutron}.

The evidence of eccentricity in this NSBH system is particularly exciting, as it offers valuable clues about the formation pathways of NSBH binaries. Understanding the formation mechanisms is one of the key objectives of GW astronomy. An isolated binary evolution channel, the currently favored channel for NSBH formation \cite{Belczynski_2002,Broekgaarden_2021,Mandel_2021}, will lead to NSBH systems with low spins and nearly circular orbits \cite{Broekgaarden_2021}. 
Therefore, the observation of measurable eccentricity in this event indicates the possibility of alternative formation processes. One such possibility involves interactions with a tertiary companion \cite{Silsbee_2017,toonen2016evolutionhierarchicaltriplestarsystems,Antonini_2017,Fragione_2019_1}, which can sustain eccentricity through angular momentum exchange between the inner binary and the outer object \cite{1910AN....183..345V,Lidov:1962wjn,1962AJ.....67..591K}. Eccentric orbits can also arise in systems assembled dynamically through interactions in dense stellar environments \cite{Freire_2004,Fragione_2019,Ye_2019,Rastello_2020,Trani_2022}. The detection of eccentricity in this NSBH candidate suggests that formation through dynamical interactions may be more common than previously anticipated. Additionally, this finding is likely to motivate renewed scrutiny of both past and future GW events for potential signatures of orbital eccentricity.

In this work, we analyze GW200105 using state-of-the-art effective-one-body (EOB) waveform models \cite{PhysRevD.59.084006,PhysRevD.62.064015,PhysRevD.62.084011,PhysRevD.64.124013,ramosbuades2023seobnrv5phmgenerationaccurateefficient,gamboa2024accuratewaveformseccentricalignedspin,PhysRevD.110.024031}, including a physically complete IMR model that incorporates both orbital eccentricity and spin precession. We examine how different choices of eccentricity priors affect the resulting eccentricity posterior distribution. In addition, we study the structure of the inferred eccentricity posterior distributions and investigate possible sources of systematics. To complement our analysis, we also utilize numerical relativity (NR) simulations to validate and deepen our understanding of the event.

The paper is organized as follows: Sec.~\ref {sec:Methods} presents an overview of the Bayesian inference framework and describes the NR code used in our study.
In Sec.~\ref{sec:results}, we present our parameter estimation results, compare them with previous studies, and assess the impact of potential sources of systematics on posterior distributions. This section also includes a discussion of our NR simulations for the event. Lastly, Sec.~\ref{sec:conclude} provides a summary of our main conclusions.

\section{Methods}
\label{sec:Methods}

\subsection{Bayesian inference}
To extract the physical properties of a compact binary system from GW observations, the data $d$ are compared against theoretical predictions of GW strain under some hypothesis $\mathcal{H}$. These predictions are generated using waveform models, which typically serve as approximations to full general relativity. The posterior distribution for the model parameters $\bm{\Theta}$ is then obtained via Bayes' theorem
\begin{equation}
    p(\bm{\Theta} |d,\mathcal{H} ) = \frac{\mathcal{L}(d|\bm{\Theta},\mathcal{H}) \pi(\bm{\Theta}|\mathcal{H})}{Z(d|\mathcal{H} )}~,
\label{eqn:bayes_theorem}
\end{equation}
where $\mathcal{L}(d|\bm{\Theta},\mathcal{H})$ denotes the likelihood function, $\pi(\bm{\Theta}|\mathcal{H})$ is the prior probability distribution and $Z(d|\mathcal{H})$, called the evidence, is calculated as $\int\mathcal{L}(d|\bm{\Theta},\mathcal{H}) \pi(\bm{\Theta}|\mathcal{H}) d \bm{\Theta}$. The evidence acts both as a normalization constant for the posterior distribution and as a means for hypothesis comparison. The relative support for two competing hypotheses, $\mathcal{H}_1$ and $\mathcal{H}_2$ can be quantified by the Bayes' factor (BF), defined as the ratio of their evidence:
\begin{equation}
    \ln{\text{BF}} = \ln \frac{Z(d|\mathcal{H}_2)} {Z(d|\mathcal{H}_1)}~.
\label{eqn:bayes_factor}
\end{equation}
Furthermore, assuming stationary, Gaussian noise in each detector, the joint log-likelihood for observing a signal $h(\bm{\Theta})$ across $N$ detectors is given by:
\begin{equation}
    \ln \mathcal{L} = \sum_{k=1}^N \langle d_k|h_k\rangle_k-\frac{1}{2}\langle d_k|d_k\rangle_k-\frac{1}{2}\langle h_k|h_k\rangle_k.
\label{eqn:lnL}
\end{equation}
The angle bracket represents the noise-weighted inner product, defined as
\begin{equation}
    \langle a |b\rangle_k = 2 \int_{|f|\geq f_{\text{low}}}^{|f|\leq f_{\text{high}}} \frac{\tilde{a}(f)^*\tilde {b}(f)}{S_k(|f|)} df~,
\label{eqn:inner_product}
\end{equation}
where $S_k(|f|)$ is the power spectral density (PSD) of the $k^{th}$ detector, and $\tilde{a}(f)$ and $\tilde{b}(f)$ are the Fourier transforms of $a(t)$ and $b(t)$. In our analysis, we evaluate the inner product over the frequency range $f_{\text{low}} = 20$ Hz to $f_{\text{high}} = 4096$ Hz, unless stated otherwise. The PSDs used for the LIGO Livingston and Virgo detectors are identical to those employed in the published LVK analysis of GW200105. Additionally, we analyze a $32$ s segment of calibrated strain data from the LIGO Livingston and Virgo detectors, sampled at $16384$ Hz.

To perform Bayesian inference, we use the {\tt RIFT} algorithm \cite{PhysRevD.92.023002,lange2018rapidaccurateparameterinference,wofford2023expandingriftimprovingperformance,wagner2025narrowingriftfocusedsimulationbasedinference}, which is well-suited for long-duration signals and waveform models with high computational cost due to its scalable, two-stage iterative approach. In the first stage, the algorithm generates GW modes $h_{l m}(\bm{\lambda})$ over a grid in the intrinsic parameter space $\bm{\lambda}$, which includes component masses $m_{1,2}$, spin vectors ${\bm{\chi}_{1,2}}$, eccentricity $e$ and the anomaly parameter $\ell$.  These modes are then used to prefilter the data, allowing rapid sampling of the extrinsic parameter space $\bm{\theta}$, independent of waveform length or evaluation time. The extrinsic parameters include luminosity distance $D_L$, inclination $\iota$, coalescence phase $\phi_c$, coalescence time $t_c$, polarization angle $\psi$, right ascension, and declination.  The likelihood is then marginalized over all extrinsic parameters to compute $\ln\mathcal{L}_{\text{marg}}$. In the second stage, $\lnLmarg(\bm{\lambda})$ values are interpolated, and adaptive importance sampling is used to construct the posterior distribution over the intrinsic parameters. This iterative process continues until convergence, at which point the posterior distribution for the extrinsic parameters is also generated. In this work, sampling in both stages is carried out using the {\tt AV} sampler (based on the Varaha \cite{Tiwari_2023,tiwari2025varahapromisingsamplerobtaining} sampler).

For waveform generation during inference, we employ models based on the EOB formalism: {\tt SEOBNRv5PHM} \cite{ramosbuades2023seobnrv5phmgenerationaccurateefficient}, {\tt SEOBNRv5EHM} \cite{gamboa2024accuratewaveformseccentricalignedspin}, and {\tt TEOBResumS-Dal\'{\i}}\footnote{The version used in this study is available at \href{https://bitbucket.org/teobresums/teobresums/src/GIOTTO/}{https://bitbucket.org/teobresums/teobresums/src/GIOTTO/} and is identified by the tag 2505.21612.} \cite{PhysRevD.110.024031} (hereafter referred to as {\tt TEOBResumS}). {\tt SEOBNRv5PHM} models waveforms from quasicircular binary black hole (BBH) systems exhibiting precession, while {\tt SEOBNRv5EHM} models eccentric BBH systems with spin vectors aligned with the orbital angular momentum. {\tt TEOBResumS} can generate waveforms from systems exhibiting both eccentricity and precession. All three models provide full IMR waveforms. In our analysis, we include all available GW modes with $l \leq 4$. The reported posterior distributions for $e$ follow the native definitions of each model. Applying {\tt gw\_eccentricity} \cite{Shaikh_2023,shaikh2025definingeccentricityspinprecessingbinaries} to our results yields nearly identical $e$ posterior distributions.  Furthermore, we define $e$ at a reference GW frequency of $20$ Hz, denoted $e_{20}$. 

In constructing the posterior distributions, we adopt uniform priors for $m_{1,2}$, $\ell$, $\psi$, $\phi_c$, and $t_c$. The sky location prior is taken to be isotropic over the celestial sphere, with the prior for $D_L$ being $\propto D_L^2$. For $\iota$, the prior is uniform in $\cos{\iota}$. Our default prior on $e$ is uniform; however, to assess the sensitivity to prior choice, we also consider a log-uniform prior.  The upper bound for the log-uniform prior is fixed at 0.2, while the lower bound is set to either $10^{-2}$ or $10^{-4}$. In aligned spin analyses, spin priors are obtained by projecting a uniform, isotropic spin distribution onto the direction perpendicular to the orbital plane. For precessing-spin analyses, we use uniform priors on spin magnitudes  $|\bm{\chi}_{i}|$ and isotropic priors on spin orientations.

When presenting some of our results, we also reformulate the mass and spin parameters into more physically meaningful combinations commonly used in GW data analysis such as the chirp mass $\mc_c = {(m_1m_2)^{3/5}/(m_1+m_2)^{1/5}}$, effective aligned spin parameter $\chi_{\text{eff}}= {(m_1\chi_{1z} + m_2\chi_{2z})/(m_1+m_2)}$ and effective precession spin parameter \cite{Schmidt_2015} $\chi_p = \max \left( \chi_{1\perp}, \frac{4q + 3}{4 + 3q} \chi_{2\perp} \right)$, where $\chi_{i\perp}$ is the in-plane spin component. 

\subsection{Numerical relativity}
We perform two sets of NR simulations of the NSBH merger: one with parameters close to those reported in \cite{morras2025orbitaleccentricityneutronstar} (labeled {NRq0.13}), and another close to the values inferred from our analysis using the {\tt TEOBResumS} model (labeled {NRq0.24}). The simulation parameters are listed in Table~\ref{tab:bhns_parameters}. For each set, we simulate both eccentric and quasicircular configurations.
The simulations are carried out using the \texttt{MAYA} code~\cite{2015ApJLEvans,2016PRDClark,2016CQGJani}, evolving the spacetime via the BSSN-Chi formulation~\cite{Shapiro1999,Shibata1995,Beyer:2004sv} with the moving puncture gauge~\cite{Campanelli2005,Baker2006}. Relativistic hydrodynamics are handled by the \texttt{Whisky} code~\cite{Baiotti:2004wn,Hawke:2005zw,Baiotti:2010zf}, and initial data are generated using a modified Bowen-York method~\cite{Clark:2016ppe} that includes matter sources. To maintain consistent resolution across all runs, we follow the criteria from \cite{Shibata:2011jka}, requiring the grid spacing $\Delta$ to satisfy $\Delta \leq M_h/20$ inside the black hole of Christodoulou mass $M_h$ and $\Delta \leq  R_*/40$ inside the neutron star of radius $R_*$. Since the black hole and the neutron star are of comparable size for {NRq0.13}, we employ eight levels of refinement with the finest grid spacing of approximately $M/86$, while {NRq0.24} adds an additional level for the black hole with finest grid spacing of $M/120$, where $M$ denotes the total mass of the system in our NR code units.

For quasicircular setups, the momenta of the boosted point sources are computed using 3PN-order quasicircular evolution \cite{Healy_2017,Ramos_Buades_2019}. To introduce eccentricity, we first use the \texttt{SEOBNRv5EHM} model to evolve the Keplerian eccentricity of $0.145$ and $0.11$ from an initial frequency of $20$ Hz to the initial separations of $D = 12.5M$ for {NRq0.13}, and $11.0M$ for {NRq0.24}, corresponding to a $(2,2)$ mode frequency of approximately $113$ Hz and $160$ Hz, respectively. The resulting eccentricity values are then used to compute a Newtonian-level correction to the tangential momentum of each point source, following the procedure described in \cite{gayathri2022eccentricityestimateblackhole}, which adjusts the quasicircular momentum to produce the target eccentric momentum.
The resulting eccentricities are verified using {\tt gw\_eccentricity}. For the quasicircular cases, {\tt gw\_eccentricity} reports a residual eccentricity of $\mathcal{O}(10^{-3})$ at the start of the waveform. All eccentricity estimates are obtained using the amplitude-based method implemented in the code.

\begin{table}
    \centering
    \addtolength{\tabcolsep}{3pt} 
    \begin{tabular}{c c c}
   \hline
   Simulations & NRq0.13 & NRq0.24 \\
    \hline
    \hline
    $M_\text{tot}/M_\odot$            & 13.0  & 11.0\\
    $q$                         & 0.13 & 0.24 \\
    Initial separation $D$ [$M$]           & 12.5 & 11.0\\
    Initial eccentricity $e_{D}$               & 0.0247 & 0.0122 \\
    \hline
  \end{tabular}
  \caption{NSBH simulation parameters: This table summarizes the parameters used for the NSBH simulation. Initial eccentricity is defined at the initial separation of the NR runs. {NRq0.13} runs use parameters close to values reported in \cite{morras2025orbitaleccentricityneutronstar} and {NRq0.24} runs use parameters close to values inferred by the {\tt TEOBResumS} model in our analysis. In both cases, the compact objects are nonspinning. The neutron star compactness, $C=m_2/R_* \approx 0.19$ for all runs.}
  \label{tab:bhns_parameters}
\end{table}

\section{Results}
\label{sec:results}
In this section, we present the results from our analysis of GW200105. In Sec.~\ref{ssec:PE_results}, we discuss the parameter estimation results obtained under different hypotheses and compare them with those reported in previous studies. In Sec.~\ref{ssec:systematics}, we explore potential sources of systematics and assess their impact on our results. In Sec.~\ref{ssec:NR_info}, we discuss the results from NR simulations.

\subsection{Parameter estimation}
\label{ssec:PE_results}
\begin{figure*}
    \includegraphics[scale=0.52]{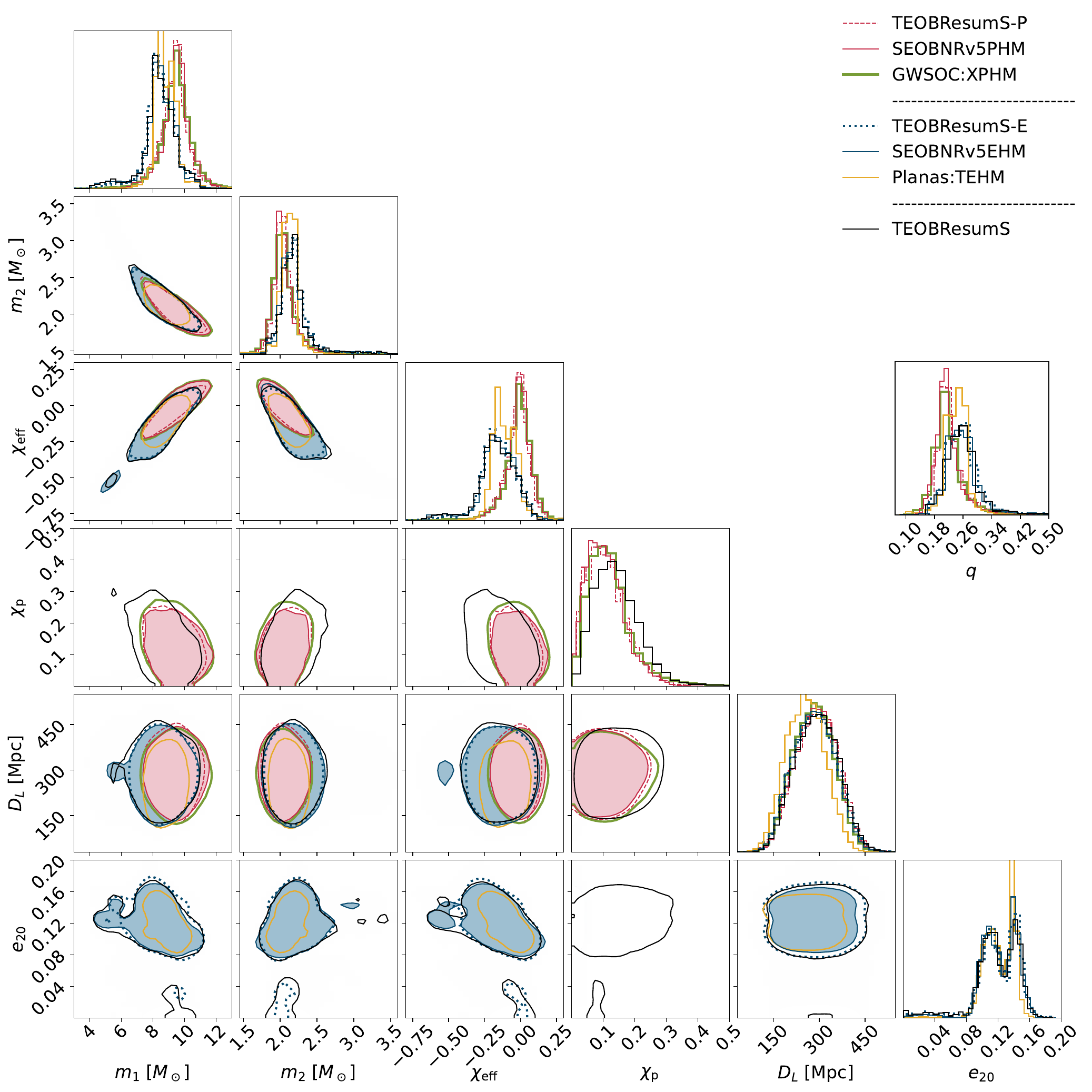}
    \caption{Comparative corner plot for GW200105. This figure shows one- and two-dimensional marginal posterior distributions for $m_{1,2}$, $\chi_\text{eff}$, $\chi_{p}$, $D_L$ and $e_{20}$. Diagonal panels show the one-dimensional marginal posterior distributions, while contours in the off diagonal panels show the 90\% credible intervals for the joint two-dimensional marginal posterior distributions.  An additional panel illustrates the marginal posterior distributions for $q$. Different colors and linestyles represent results from various waveform models, grouped according to three distinct hypotheses for clearer comparison. Two publicly available analysis results are also included for comparison.}
    \label{fig:main_corner}
\end{figure*}

We carry out Bayesian inference on GW200105 under three distinct hypotheses: (1) the binary is quasicircular and precessing, (2) it is eccentric with spins aligned to the orbital angular momentum, and (3) it is eccentric as well as precessing. Figure~\ref{fig:main_corner} presents the posterior distributions for several key parameters obtained under all three hypotheses. For comparison, we also include posterior distributions from previous studies \cite{Abbott_2021,planas2025eccentricinspiralmergerringdownanalysisneutron}. Structuring our analysis around these separate hypotheses enables a direct comparison with earlier work, each of which carried out inference under one of these hypotheses. It also allows us to compare the results across different hypotheses within a consistent framework.

\begin{table*}
\centering
\addtolength{\tabcolsep}{3pt} 
     \begin{tabular}{c c c c c c c c c c c}
     \hline
      Model & $m_1/\text{M}_\odot$ & $m_2/\text{M}_\odot$  & $M_{\text{tot}}/\text{M}_\odot$  & $\mc_c/\text{M}_\odot$ & $q$ & $\chi_{\text{eff}}$ & $\chi_{\text{p}}$ & $e_{20}$ & $D_L/\text{Mpc}$  \\[0.5ex] 
     \hline\hline
     {\tt TEOBResumS-P} & $9.42^{+1.27}_{-1.70}$ & $2.04^{+0.34}_{-0.18}$ & $11.46^{+1.08}_{-1.37}$ & $3.62^{+0.01}_{-0.01}$ &
     $0.22^{+0.09}_{-0.04}$ & $-0.01^{+0.11}_{-0.17}$ & $0.10^{+0.13}_{-0.08}$ & ... & $291^{+112}_{-114}$  
     \\[0.75ex]
    {\tt SEOBNRv5PHM} & $9.52^{+1.40}_{-1.64}$ & $2.02^{+0.32}_{-0.20}$ & $11.54^{+1.21}_{-1.32}$ & $3.62^{+0.01}_{-0.01}$ &
     $0.21^{+0.08}_{-0.04}$ & $-0.01^{+0.11}_{-0.16}$ & $0.10^{+0.12}_{-0.08}$ & ... & $291^{+107}_{-111}$  
     \\[0.75ex]
     {\tt GWSOC:XPHM} \cite{Abbott_2021}& $9.48^{+1.55}_{-1.77}$ & $2.03^{+0.35}_{-0.21}$ & $11.51^{+1.33}_{-1.41}$ & $3.62^{+0.01}_{-0.01}$ &
     $0.21^{+0.09}_{-0.05}$ & $-0.01^{+0.12}_{-0.18}$ & $0.11^{+0.15}_{-0.08}$ & ... & $281^{+107}_{-108}$  
     \\[1.5ex]

     {\tt TEOBResumS-E} & $8.44^{+1.59}_{-1.74}$ & $2.17^{+0.45}_{-0.24}$ & $10.61^{+1.36}_{-1.28}$ & $3.58^{+0.03}_{-0.04}$ & $0.26^{+0.13}_{-0.06}$ & $-0.16^{+0.19}_{-0.20}$ & ... & $0.12^{+0.03}_{-0.06}$ & $285^{+116}_{-113}$ 
     \\[0.75ex]
     {\tt SEOBNRv5EHM} & $8.47^{+1.54}_{-2.54}$ & $2.16^{+0.72}_{-0.23}$ & $10.64^{+1.30}_{-1.82}$ & $3.58^{+0.03}_{-0.04}$ & $0.26^{+0.23}_{-0.06}$ & $-0.15^{+0.19}_{-0.31}$ & ... & $0.12^{+0.03}_{-0.03}$ & $282^{+111}_{-111}$
     \\[0.75ex]
     {\tt Planas:TEHM} \cite{planas2025eccentricinspiralmergerringdownanalysisneutron}&$8.71^{+1.25}_{-0.75}$ & $2.13^{+0.15}_{-0.18}$ & $10.83^{+1.06}_{-0.60}$ & $3.58^{+0.03}_{-0.02}$ & $0.24^{+0.04}_{-0.05}$ & $-0.12^{+0.15}_{-0.11}$ & ... & $0.12^{+0.02}_{-0.03}$ & $255^{+104}_{-102}$ 
     \\[1.5ex]
      
     {\tt TEOBResumS} & $8.45^{+1.67}_{-2.46}$ & $2.17^{+0.70}_{-0.26}$ & $10.61^{+1.42}_{-1.74}$ & $3.58^{+0.03}_{-0.04}$ & $0.26^{+0.22}_{-0.07}$ & $-0.16^{+0.19}_{-0.28}$ & $0.14^{+0.14}_{-0.09}$ & $0.12^{+0.03}_{-0.07}$ & $293^{+113}_{-114}$
     \\[0.75ex]
     {\tt Morras:pyEFPE} \cite{morras2025orbitaleccentricityneutronstar} & $12.25^{+0.70}_{-1.96}$ & $1.61^{+0.27}_{-0.05}$ & $13.86^{+0.67}_{-1.69}$ & $3.54^{+0.07}_{-0.02}$ & $0.13^{+0.05}_{-0.01}$ & $0.00^{+0.10}_{-0.09}$ & $0.06^{+0.13}_{-0.04}$ & $0.145^{+0.007}_{-0.09}$ & $296^{+108}_{-124}$ 
     \\[0.75ex]
     
     \hline \\
     \end{tabular}
     \caption{Summary statistics for GW200105. This table reports median values along with the 90\% credible intervals obtained from the analysis of GW200105 under different hypotheses. For comparison, results from previous studies are also included. All models that incorporate eccentricity, whether eccentric only or including precession, consistently show evidence of eccentricity in the signal.}
    \label{tab:median_values}
\end{table*}

\subsubsection{Precessing only}
Under our first hypothesis, we perform parameter estimation using {\tt SEOBNRv5PHM} and {\tt TEOBResumS}. In the case of {\tt TEOBResumS}, we fix the $e$ and $\ell$ to zero, and refer to this configuration as {\tt TEOBResumS-P}. This setup involves inference over a 15-dimensional parameter space. We compare the results obtained using these two models with those made publicly available by the LVK Collaboration \cite{Abbott_2021}. To explore the effects of waveform modeling systematics, we specifically include posterior distributions obtained using {\tt IMRPhenomXPHM} \cite{Pratten_2021}, a member of the phenomenological waveform family \cite{Ajith_2007}. The comparison reveals excellent agreement across all three models: the posterior distributions for key parameters, shown in Fig.~\ref{fig:main_corner}, are nearly indistinguishable, and the 90\% credible summary statistics in Table~\ref{tab:median_values} reinforce this agreement. Our analysis of GW200105 under this hypothesis supports previous findings, specifically the system has $\mc_c \approx 3.62$, $q \approx 0.21$, and $\chi_{\text{eff}}$ centered near zero, with comparable support for both positive and negative values. Additionally, we find $\chi_{\text{p}} \approx 0.11$, suggesting weak evidence for precession. In both analyses, the maximum $\lnLmarg$ value is approximately 76.

\subsubsection{Eccentric only}
\begin{figure}
    \includegraphics[scale=0.38]{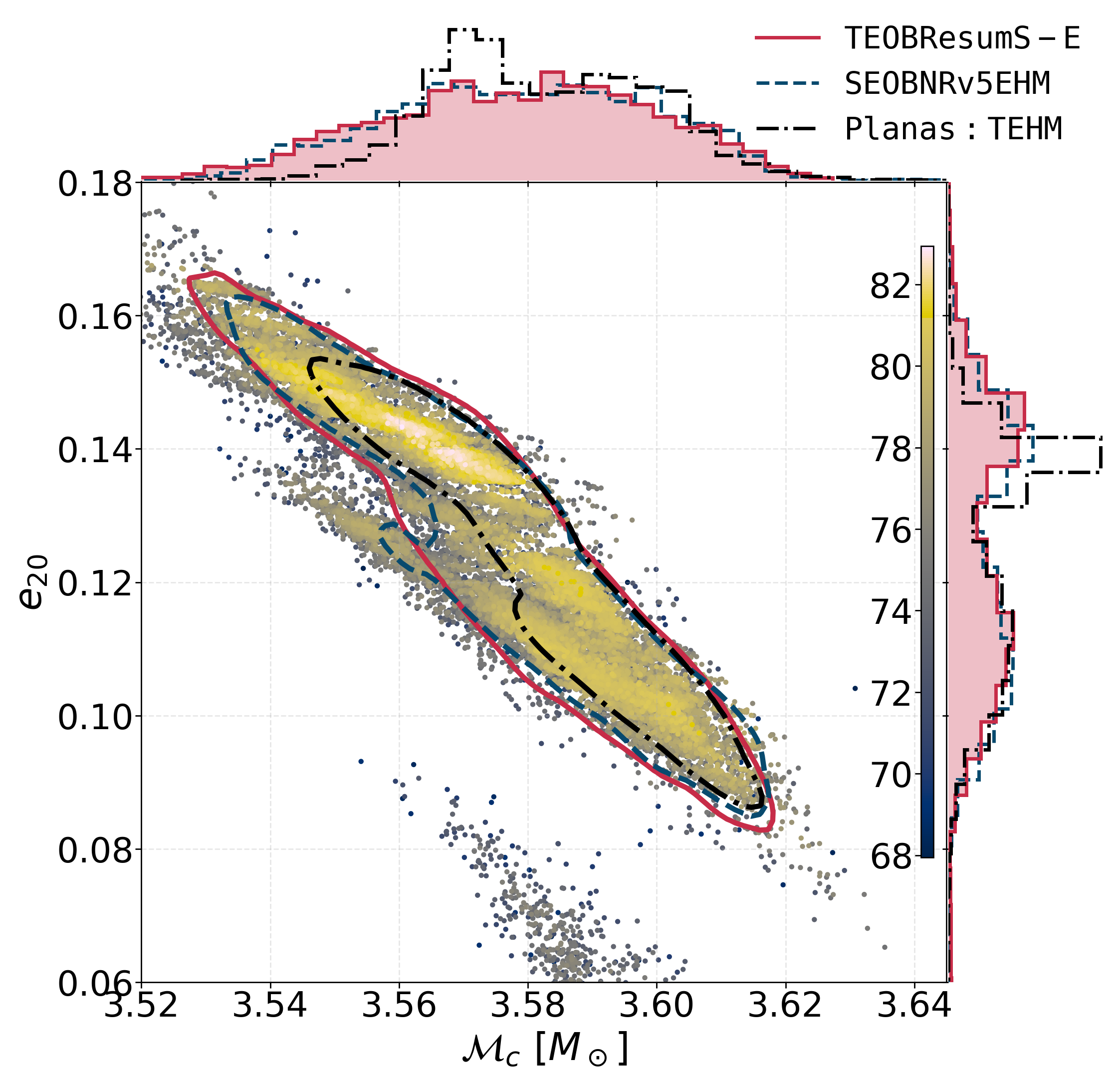}
    \caption{Comparison of $\mc_c$ and $e_{20}$ distributions from eccentric-only hypothesis. This figure shows the marginal one- and two-dimensional posterior distributions for $\mc_c$ and $e_{20}$ obtained from {\tt TEOBResumS-E} and {\tt SEOBNRv5EHM} analyses, with {\tt Planas:TEHM} included for comparison. The two-dimensional contours represent the $90\%$ credible interval. The scatter plot uses a color gradient to highlight points that lie within $15$ of the maximum $\lnLmarg$, based on the {\tt TEOBResumS-E} results. For clarity, a reduced eccentricity range is shown.}
    \label{fig:eccentricity_comparison}
\end{figure}

Under our second hypothesis, we perform parameter estimation using {\tt SEOBNRv5EHM} and {\tt TEOBResumS}. In the latter case, transverse spin components are set to zero, a configuration we denote as {\tt TEOBResumS-E}. This setup involves inference over a 13-dimensional parameter space. Under this hypothesis, both waveform models provide significantly better fits to the data compared to their precession-only counterparts, with maximum $\lnLmarg$ values reaching $\approx 82.8$, a gain of about $7$. To quantify the preference for eccentricity, we compute the $\ln \text{BF}$ between the eccentric-only and precession-only models, finding a value of about $2.5$ for the two {\tt SEOBNRv5} approximants and $3.8$ for the two {\tt TEOBResumS} configurations, moderately favoring the eccentric hypothesis. We also observe structure in the likelihood surface, which is particularly reflected in the posterior distributions for intrinsic parameters, as shown in Fig.~\ref{fig:main_corner} and Fig.~\ref{fig:eccentricity_comparison}.  Additionally, the median chirp mass $\mc_c$ inferred under the eccentric-only hypothesis falls outside the $90 \%$ credible interval of the precession-only result. This is consistent with the known negative correlation between eccentricity and $\mc_c$: neglecting eccentricity can bias the chirp mass estimate upward when the true signal is eccentric \cite{PhysRevD.105.023003}. Incorporating eccentricity also leads to a shift in the median $\chi_{\text{eff}}$ to approximately $-0.15$, along with an increased support for negative $\chi_{\text{eff}}$ values. We also observe a decrease in the inferred detector-frame black hole mass and a corresponding increase in the neutron star mass. A closer examination of the $\lnLmarg$ values shows that the secondary peak in $e_{20}$ is associated with a higher black hole mass, a lower neutron star mass, and a value of $\chi_{\text{eff}}$ closer to zero. Finally, for both waveform models, the maximum likelihood parameters are: $\mc_c \approx 3.57\ \text{M}_\odot$, $q \approx 0.27$, $\chi_{\text{eff}} \approx -0.2$, and $e_{20} \approx 0.14$.

As summarized in Table~\ref{tab:median_values}, the results obtained from both models are consistent with each other. Figure~\ref{fig:main_corner} further illustrates this agreement, showing that the posterior distributions from the two models almost completely overlap. Importantly, both analyses consistently support the presence of nonzero eccentricity in the observed signal. We also compare our results with those from \cite{planas2025eccentricinspiralmergerringdownanalysisneutron}, which used the {\tt IMRPhenomTEHM} \cite{planas2025timedomainphenomenologicalmultipolarwaveforms} for inference. Across all three eccentric-only models, we find good agreement in the inferred mass and spin parameters. Additionally, all analyses reveal a multimodal structure in the eccentricity posterior distribution. However, we find that the {\tt IMRPhenomTEHM} analysis yields tighter constraints on the parameters compared to our results, as shown in both Table~\ref{tab:median_values} and Fig.~\ref{fig:eccentricity_comparison}.

\subsubsection{Eccentric and precessing}
\label{sssec:e+p}
For our third hypothesis, we perform parameter estimation using {\tt TEOBResumS}, exploring a 17-dimensional parameter space. The resulting posterior distributions are in strong agreement with those obtained under the eccentric-only hypothesis, as shown in Fig.~\ref{fig:main_corner} and in Table~\ref{tab:median_values}. In particular, the posterior distributions obtained with {\tt TEOBResumS} closely match those from {\tt TEOBResumS-E}, indicating that the inclusion of precession has minimal impact on the recovered parameters for this event. There is a slight improvement in $\lnLmarg$ when precession is included, with it reaching $83.4$, a gain of around $0.6$. 
Additionally, the $\chi_{\text{p}}$ posterior distribution is consistent with those from the precession-only analyses, reinforcing the conclusion that the system exhibits only weak precession.

\begin{figure}
    \includegraphics[scale=0.37]{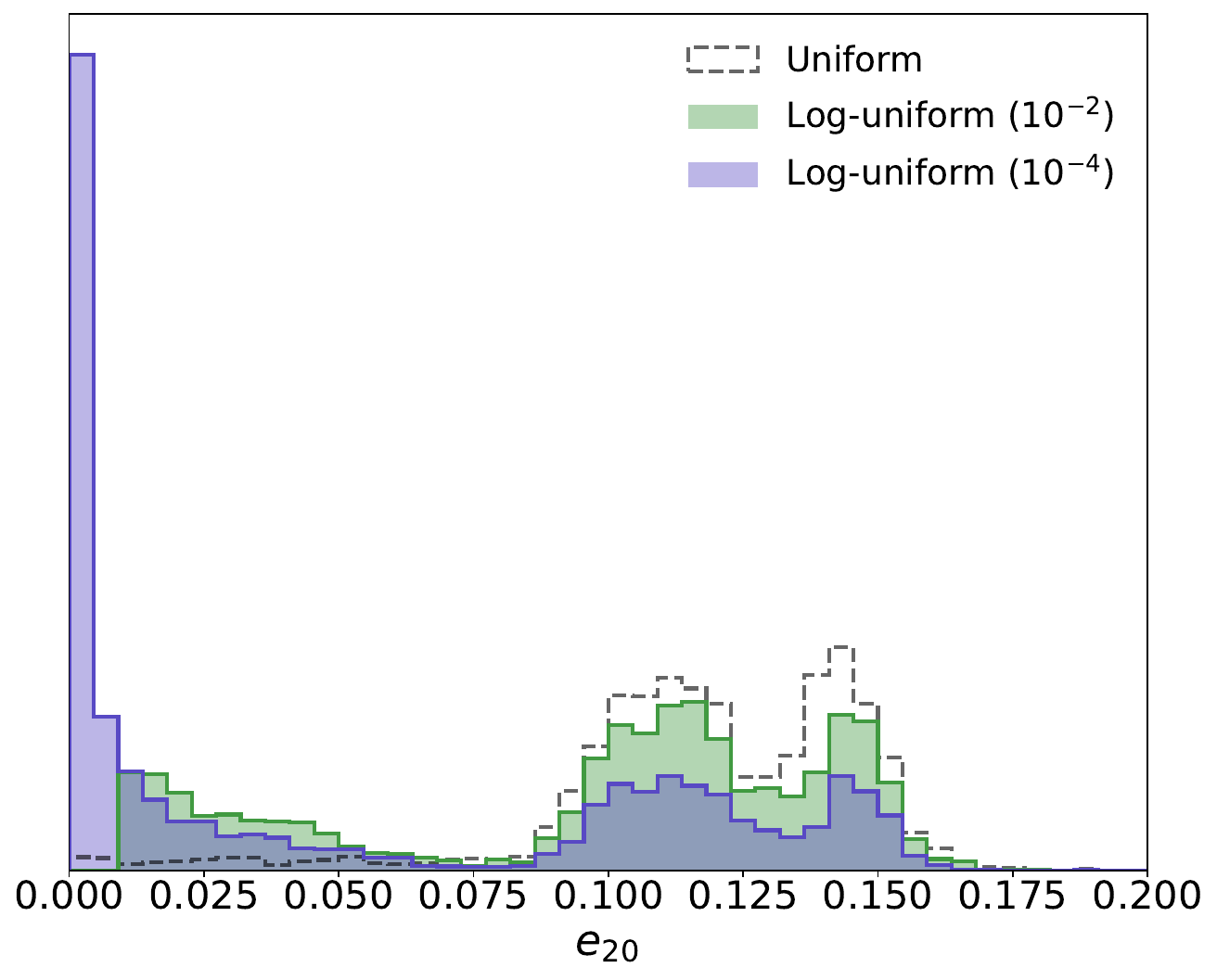}
    \caption{Impact of eccentricity prior. This figure compares the $e_{20}$ posterior distributions obtained under the eccentric-precessing hypothesis with {\tt TEOBResumS}, using both the uniform and log-uniform priors. The log-uniform prior is evaluated with two different lower bounds: $10^{-2}$ and $10^{-4}$. Even in the more extreme case, where the lower bound is $10^{-4}$, the posterior retains substantial support at nonzero $e_{20}$ values.}
    \label{fig:eccentricity_prior_comparison}
\end{figure}

We also perform inference using a log-uniform prior on $e_{20}$ and compare the resulting posterior distributions with those obtained under a uniform prior, as shown in Fig.~\ref{fig:eccentricity_prior_comparison}. Since log is undefined at zero, we set nonzero lower bounds for the log-uniform prior, specifically $10^{-2}$ and $10^{-4}$ in separate runs. A log-uniform prior assigns equal weight to each order of magnitude and inherently downweights larger values. As seen in Fig.~\ref{fig:eccentricity_prior_comparison}, decreasing the lower bound leads to a suppression of posterior weight at higher eccentricities. Nonetheless, even with the more aggressive prior favoring low eccentricity, a nontrivial amount of posterior support remains at larger $e_{20}$ values.

\subsection{Investigation of possible systematics}
\label{ssec:systematics}
In this section, we explore potential sources of systematics that could impact the parameter estimation results presented in Sec.~\ref{ssec:PE_results}. The parameter estimation settings used in our main analysis were chosen to closely match those adopted in the original LVK study \cite{Abbott_2021}, and using those settings, we successfully reproduce their precession-only results. When eccentricity is included in the analysis, we observe an increase in the likelihood of the best-fit waveform. However, two key issues remain to be addressed:
\begin{enumerate}
    \item The discrepancy between our results and those reported in \cite{morras2025orbitaleccentricityneutronstar}.
    \item The origin of the multimodal structure observed in the $e_{20}$ posterior distribution.
\end{enumerate}
In \cite{morras2025orbitaleccentricityneutronstar}, the $e_{20}$ posterior showed no evidence of multimodality, and the authors reported $M_\text{tot} = 13.86^{+0.67}_{-1.69}$ and $q = 0.13^{+0.05}_{-0.01}$. In contrast, our analysis using the {\tt TEOBResumS} waveform model yields $M_\text{tot} = 10.61^{+1.42}_{-1.74}$ and $q = 0.26^{+0.22}_{-0.07}$. The results for mass, aligned spin, and eccentricity parameters obtained from both our eccentric-only and combined eccentricity-precession analyses show good agreement with the findings of \cite{planas2025eccentricinspiralmergerringdownanalysisneutron}, which also utilized a full IMR waveform model with higher-order modes, though without incorporating precession. Motivated by these differences, we now carry out a targeted investigation to understand the potential origins of these differences.

We begin by conducting parameter estimation runs using settings intended to match those employed in \cite{morras2025orbitaleccentricityneutronstar}. This required adjusting the upper frequency cutoff in the inner production computation defined in Eq.\eqref{eqn:inner_product}. In our main analysis, we use an upper limit of $4096$ Hz. In contrast, their analysis, based on an inspiral-only model, limited the frequency integration to $280$ Hz. Accordingly, we adopt a $f_\text{max}$ of 280 Hz in our reanalysis. Additionally, to match their GW mode content, we restrict the model to include only the $(2,\pm2)$ modes. 
Using these modified settings, we reanalyzed the event with {\tt TEOBResumS}, exploring the full $17$-dimensional parameter space including both eccentricity and precession. 

\begin{figure}
    \includegraphics[scale=0.37]{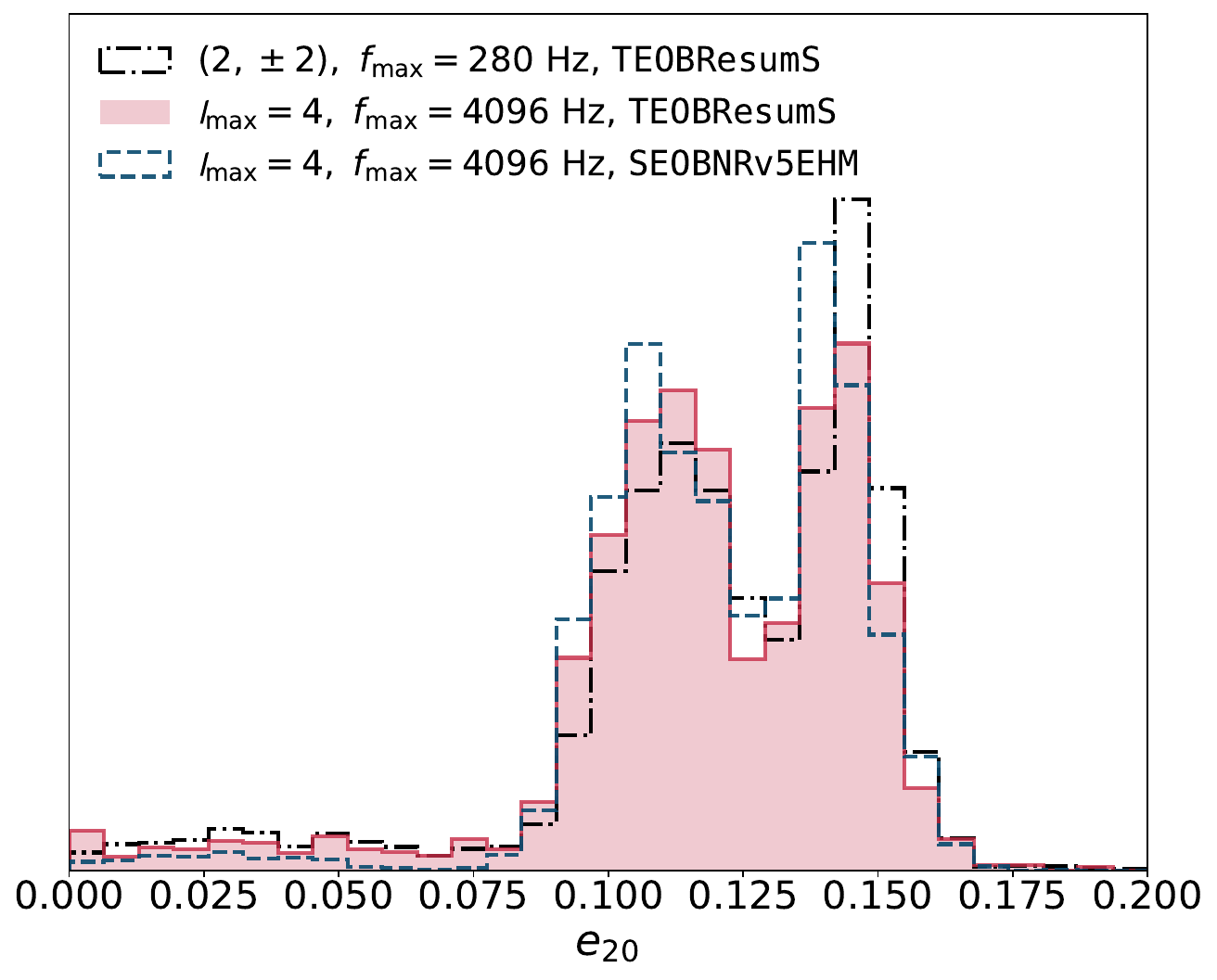}
    \caption{Systematics study. This figure shows the $e_{20}$ posterior distribution obtained using parameter estimation settings matched to those employed in \cite{morras2025orbitaleccentricityneutronstar}. For comparison, results from the main analyses are also shown. Despite similar settings, the multimodal structure persists.}
    \label{fig:systematics_study}
\end{figure}

Under these modified settings, the difference in the maximum $\lnLmarg$ between the two $e_{20}$ modes experiences a slight increase; however, the overall multimodal structure remains, as shown in Fig.~\ref{fig:systematics_study}.  We also continue to recover $q$ close to the value inferred under our standard settings, around $0.26$. Due to the lower $f_\text{max}$, maximum $\lnLmarg$ decreases from $83.4$ to $79.0$. These results indicate that even when adopting settings similar to those in \cite{morras2025orbitaleccentricityneutronstar}, our findings remain broadly consistent with the main results in Sec.\ref{ssec:PE_results}, suggesting that waveform model systematics likely accounts for the observed differences.

We also investigate the impact of segment length on our parameter estimation results. Since GW data analysis is carried out in the frequency domain, and the frequency resolution in that domain is inversely proportional to the segment length, this choice can potentially impact the results. As noted in \cite{planas2025eccentricinspiralmergerringdownanalysisneutron}, a $32$ s segment length limits the frequency resolution below $30$ Hz. Our main analysis uses the $32$ s segment length, in line with the original LVK settings. To explore the potential impact of this choice, we carry out a zero-noise injection-recovery using {\tt TEOBResumS}, with both 32 s and 128 s segment lengths. The injection corresponds to the maximum $\text{ln}\mathcal{L}$ point from our eccentric-precessing analysis of the real event. We find that the recovered $e_{20}$ posterior distributions from these two zero-noise runs exhibit no multimodality and are nearly identical, ruling out the segment length as the source of the observed multimodal structure. The absence of multimodality in our zero-noise injection analyses is consistent with the findings of \cite{planas2025eccentricinspiralmergerringdownanalysisneutron}, which likewise reported no evidence of such structure when analyzing a 32 s zero-noise injection.

Finally, we reanalyze GW200105 with {\tt TEOBResumS}, this time using an alternative sampler and a different initial grid. The sampler employed is the Gaussian mixture sampler implemented in {\tt RIFT}, and the evaluation points in the initial grid are randomly generated within a region enclosing the high-$\lnLmarg$ region. The resulting posterior distributions are virtually identical to those presented in Sec.~\ref{sssec:e+p}.  Alongside the results from our earlier tests, this strongly indicates that the observed multimodality arises from the data themselves, rather than from choices of sampling or parameter estimation settings.

\subsection{Information from numerical relativity}
\label{ssec:NR_info}
While our expectation is that the model waveforms fully describe the event as measured by the detectors, NR can provide novel insights into the final fate of the merger. The NR simulations enable us to estimate the system's mass and radiated energy, and to compare these quantities for both the quasicircular and eccentric hypotheses.
The simulations were carried out at the parameters listed in Table~\ref{tab:bhns_parameters}. The listed mass ratio values fall within the regime typically associated with nondisruptive mergers~\cite{Shibata:2011jka}, where most of the neutron star is swallowed by the black hole with minimal matter left behind. This expectation is confirmed in Table~\ref{tab:nr_final_param}, where the mass remaining $M_{\text{rem}}$ outside a radius of $r = 20M$ is found to be $2$ orders of magnitude smaller than the GW-radiated energy $E_{\text{rad}}$. 

\begin{figure}
    \includegraphics[scale=0.35]{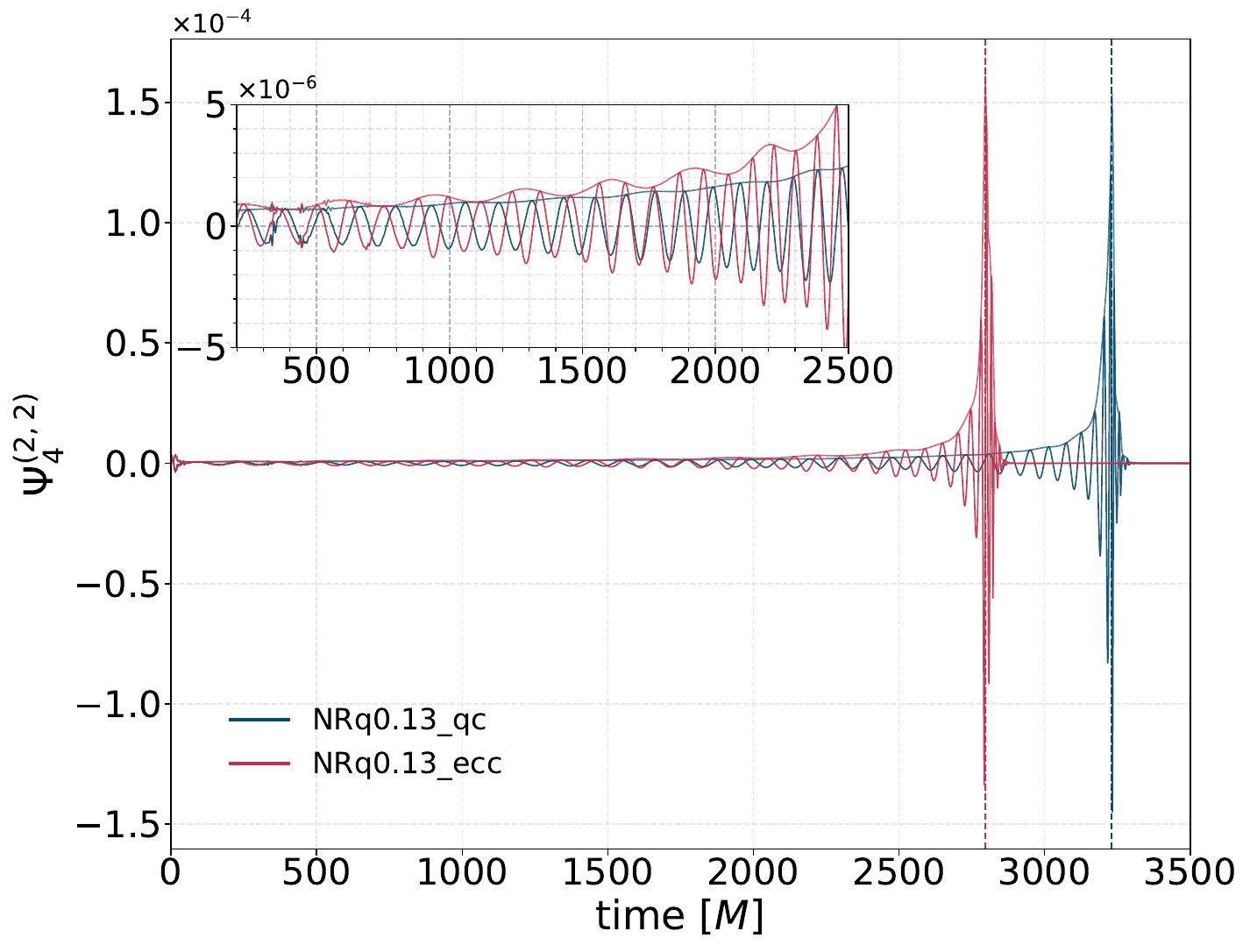}
    \caption{Gravitational wave comparison. This figure shows the $(2,2)$ mode of the Weyl scalar $\Psi_4^{(2,2)}$ extracted at a radius of $r = 130M$ for both the circular and eccentric cases of {NRq0.13}, plotted against retarded time. The dashed vertical lines at $t_{\text{qc}} = 3231M$ and $t_{\text{ecc}} = 2796M$ mark the merger times, defined by the peaks in the waveform amplitude.}
    \label{fig:nr_psi4}
\end{figure}

To estimate the merger time, we extract GWs by placing virtual detectors at a radius of $r = 130M$ and compute the Weyl scalar $\Psi_4$, as shown in Fig.~\ref{fig:nr_psi4}. The merger time is identified by the peak of the $(2,2)$ mode of $\Psi_4$. We find that the eccentric binary cases merge earlier than the circular cases for both sets of NR runs, as the merger time $(t_{\text{qc}}, t_{\text{ecc}})$ was recorded to be $(3231,2796)M$ for NRq0.13 and $(1440,1358)M$ for NRq0.24. This outcome is consistent with expectations, as binaries with higher eccentricity radiate energy more efficiently, leading to faster coalescence~\cite{Peters_1964}. 
Despite the earlier merger in the eccentric case, we find that the final black hole mass and spin, the remnant mass, and the total radiated energy are not significantly affected by the inclusion of eccentricity, consistent with the general conclusion reported in \cite{Nee_2025}. We therefore conclude that the inclusion of eccentricity in this system does not lead to significant differences in the final-state properties of the remnant black hole.

\begin{table}
    \centering
    \addtolength{\tabcolsep}{3pt} 
    \begin{tabular}{c c c c c c c}
    \hline
                     & $M_h/M$  &  $\chi_z$ &   $E_{\text{rad}}/M$ & $M_{\text{rem}}/M$\\
    \hline\hline
        NRq0.13-qc           & 0.990 & 0.323 & 7.22e-3 & 2.20e-5 \\
        NRq0.13-ecc          & 0.990 & 0.322 & 7.13e-3 & 1.83e-5 \\
        NRq0.24-qc           & 0.981 & 0.461 & 1.40e-2 & 1.01e-4 \\
        NRq0.24-ecc         & 0.981 & 0.461 & 1.40e-2 & 1.01e-4 \\
    \hline 
  \end{tabular}
  \caption{Postmerger properties of the system from NR simulations: This table summarizes the postmerger properties of the system under both quasicircular (qc) and eccentric (ecc) hypotheses, based on simulations using the mass and spin parameters listed in Table~\ref{tab:bhns_parameters}. All quantities were evaluated $50M$ after the merger.}
  \label{tab:nr_final_param}
\end{table}

\section{Conclusions}
\label{sec:conclude}
In this study, we performed a detailed analysis of GW200105 using state-of-the-art EOB waveform models. This included {\tt TEOBResumS}, which models binaries with both orbital eccentricity and spin precession across the full IMR regime, and also incorporates higher-order GW modes. This work represents the first application of a physically complete waveform model to this event. We also presented results obtained with \texttt{SEOBNRv5EHM} and \texttt{SEOBNRv5PHM}. We analyzed this event under three different hypotheses: precessing only, eccentric only, and eccentric as well as precessing. We found that introducing eccentricity in the analysis increases the maximum $\lnLmarg$ compared to the precessing-only results, indicating a better fit between the model and the observed data. However, inclusion of eccentricity parameters in the analysis increased the complexity of the $\lnLmarg$ surface, leading to a more challenging parameter estimation. Furthermore, when both eccentricity and precession are included in the analysis, the gain in $\lnLmarg$ is only marginal relative to the eccentric-only case, and the resulting posterior distributions remain largely unchanged, supporting the conclusion from the precession-only analysis that the data offer limited evidence for precession.

Across all analyses involving eccentricity, we found strong support for nonzero eccentricity. Under the eccentric-precessing hypothesis, the inferred eccentricity at $20$ Hz is $0.12^{+0.03}_{-0.07}$, with zero eccentricity excluded at the $99\%$ credible level. Notably, the posterior distribution for $e_{20}$ was multimodal. We investigated this structure in the $e_{20}$ posterior distribution by conducting targeted tests. These tests revealed no indication that the sampling or parameter estimation settings are responsible for the observed multimodality, suggesting instead that it reflects genuine features supported by the waveform models used. In addition, under the eccentric-precessing hypothesis, we infer a mass ratio of $q=0.26^{+0.22}_{-0.07}$, which differs from the value $q=0.13^{+0.05}_{-0.01}$ reported in \cite{morras2025orbitaleccentricityneutronstar}. Our investigation indicated that this discrepancy is likely due to waveform modeling systematics.

Additionally, we performed NR simulations using parameters similar to those reported in \cite{morras2025orbitaleccentricityneutronstar} and those obtained here under the eccentric-precessing hypothesis. By comparing the quasicircular and eccentric simulations, we find that including eccentricity does not significantly affect key physical quantities such as the remnant peak luminosity, total radiated energy, final black hole mass, or spin. \\

\section*{ACKNOWLEDGMENTS}
We would like to thank Harald Pfeiffer, Patricia Schmidt, Katelyn Wagner, Md Arif Shaikh and Gonzalo Morras for their helpful comments. We also thank Maria de Lluc Planas and collaborators for publicly sharing their posterior samples. We thank the anonymous reviewer for their helpful comments. The posterior samples generated in our main analyses are available at \cite{GW200105_samples}. A.J. and D.S. acknowledge support from NASA Grant No. 80NSSC24K0437, and from NSF Grants No. PHY-2207780 and  No. PHY-2114581. B.T. and P.L. acknowledge support from NSF Grants No. PHY-2114582 and No. PHY-2207780.
R.O.S.  gratefully acknowledges support from NSF Awards No. NSF PHY-1912632, No. PHY-2012057, No. PHY-2309172, No. AST-2206321, and the Simons Foundation. The computing resources necessary to perform the NR simulations were provided by TACC PHY20039. 
This material is based upon work supported by NSF's LIGO Laboratory which is a major facility fully funded by the NSF. The authors are grateful for computational resources provided by LIGO Laboratory and supported by NSF Grants PHY-0757058 and PHY-0823459.  
The work was done by members of the Weinberg Institute and has an identifier of UT-WI-38-2025‌.

\section*{DATA AVAILABILITY}
The data that support the findings of this article are openly available from the Gravitational Wave Open Science Center at \cite{GW200105_data}.

\bibliography{references}

\end{document}